\documentclass[preprint,superscriptaddress,aps]{revtex4}
\usepackage{amsfonts}
\usepackage{amssymb}
\usepackage{amsmath}
\usepackage{graphics}
\usepackage{graphicx}
\usepackage{color}
\usepackage[T1]{fontenc}
\usepackage{hyperref}
\newcommand{\bea}{\begin{eqnarray}}
\newcommand{\eea}{\end{eqnarray}}

\begin{document}
\title{On the generic higher-derivative $\mathcal{N}=2$, $d=3$ gauge theory}

\author{F. S. Gama}
\email{fgama@fisica.ufpb.br}
\affiliation{Departamento de F\'{\i}sica, Universidade Federal da Para\'{\i}ba\\
 Caixa Postal 5008, 58051-970, Jo\~ao Pessoa, Para\'{\i}ba, Brazil}

\author{M. Gomes}
\email{mgomes@fma.if.usp.br}
\affiliation{Departamento de F\'{\i}sica Matem\'atica, Universidade de S\~ao Paulo,\\
 Caixa Postal 66318, 05314-970, S\~ao Paulo, SP, Brazil}

\author{J. R. Nascimento}
\email{jroberto@fisica.ufpb.br}
\affiliation{Departamento de F\'{\i}sica, Universidade Federal da Para\'{\i}ba\\
 Caixa Postal 5008, 58051-970, Jo\~ao Pessoa, Para\'{\i}ba, Brazil}

\author{A. Yu. Petrov}
\email{petrov@fisica.ufpb.br}
\affiliation{Departamento de F\'{\i}sica, Universidade Federal da Para\'{\i}ba\\
 Caixa Postal 5008, 58051-970, Jo\~ao Pessoa, Para\'{\i}ba, Brazil}
 
\author{A. J. da Silva}
\email{ajsilva@fma.if.usp.br}
\affiliation{Departamento de F\'{\i}sica Matem\'atica, Universidade de S\~ao Paulo,\\
 Caixa Postal 66318, 05314-970, S\~ao Paulo, SP, Brazil}
 
\begin{abstract}
We formulate a generic $\mathcal{N}=2$ three-dimensional superfield higher-derivative gauge theory coupled to the matter, which, in certain cases reduces to the $\mathcal{N}=2$ three-dimensional scalar super-QED, or supersymmetric Maxwell-Chern-Simons or Chern-Simons theories with matter. For this theory, we explicitly calculate the one-loop effective potential.
\end{abstract}

\maketitle

\section{Introduction}

The idea of extending  field theory models through the introduction of additive higher-derivative terms is very old. Initially, higher derivative models have been introduced in the gravity context, aiming to construct a renormalizable gravity theory \cite{Stelle}. This gave rise to a great amount of studies on the quantum properties of different gravity models (for a review on these studies, see \cite{BO}). 

It is clear that the introduction of higher derivatives in a theory highly improves its renormalization properties possibly making it finite. Therefore it was natural that within the supersymmetry context, higher derivatives were first considered in a (super)gravity model~\cite{anomaly}. For the $\mathcal{N}=1$ supergravity, the quantum dynamics for its higher-derivative extension has been studied \cite{dilaton}, and only recently, one-loop studies for more generic examples of  higher-derivative four-dimensional supersymmetric theories have been carried out \cite{ourhigh}. In the three-dimensional case, the effective action for higher-derivative supersymmetric field theories have been studied in \cite{Our1} at the one-loop level, and some two-loop aspects of the higher-derivative $3D$ super-QED were considered in \cite{Gall}.

In this work, we extend our studies to a new class of theories -- we present the higher-derivative extension of the three-dimensional theories with extended supersymmetry. It is known that the $\mathcal{N}=2$ superfield formulation in three-dimensional space-time is very similar to the $\mathcal{N}=1$ superfield formalism for the four-dimensional space-time \cite{Hit}. Recently, studies of the $\mathcal{N}=2$ supersymmetric theories within this formulation have been carried out in \cite{Buch}.
Within this paper, we study the one-loop effective action for the higher-derivative theories described in this formalism.

Throughout this paper, we are using the conventions of \cite{Hit}, constructed on the base of those ones used in \cite{SGRS}. They are explained in the Appendix.

The structure of the paper is as follows. In the section II we present the classical formulation of the higher-derivative ${\cal N}=2$, $d=3$ supersymmetric gauge theory, in the section III we carry out  the one-loop calculations, and the section IV is a Summary where the results are discussed. In the Appendix, the main relations for the ${\cal N}=2$, $d=3$ supersymmetry algebra are given. 

\section{Higher-derivative $\mathcal{N}=2$, $d=3$ gauge theory}

In the pure gauge sector, let us start with the most general $\mathcal{N}=2$, $d=3$ abelian gauge action which will be used to find the one-loop K\"{a}hlerian effective potential (KEP)
\bea
\label{puregauge}
S_{G}=\frac{1}{2}\int d^3xd^4\theta\, V[f(\Box)\bar D^\alpha D_\alpha+h(\Box)D^\alpha \bar D^2D_\alpha]V \ ,
\eea
where $f(\Box)$ and $h(\Box)$ are analytical functions of the d'Alembertian operator $\Box$. In particular, if (up to multiplicative constants) $f=m$, where $m$ is a constant with mass dimension, and $h=0$ we have a Chern-Simons theory; if $f=0$ and $h=1$ we have a Maxwell theory, and if $f=m$ and $h=1$ we have a Maxwell-Chern-Simons theory. If $f$ and/or $h$ involves higher degrees of $\Box$ we have a higher-derivative supersymmetric gauge theory.

The structure of the expression (\ref{puregauge}) deserves some justification. Firstly, we have ignored in (\ref{puregauge}) terms higher than quadratic in the gauge superfield $V(z)$ due to the fact that the KEP is, by definition, a function only of background matter superfields, and such terms necessarily contribute with background gauge superfields in one loop. Therefore, terms higher than quadratic in $V(z)$ do not contribute to one-loop KEP. Secondly, for simplicity, we are working with an abelian theory because  the one-loop KEP for a non-Abelian theory is the same as for an Abelian one, up to the constant depending on the algebraic factor, again due to the fact that the self-coupling of the gauge superfield does not
contribute to the one-loop KEP. Lastly, $S_G$ is invariant under the gauge transformation $\delta V=i(\bar\Lambda-\Lambda)$ because the operators $\bar D^\alpha D_\alpha$ and $D^\alpha \bar D^2D_\alpha$ commute with $\Box$ and annihilate the superfields $\bar\Lambda$ and $\Lambda$ which satisfy the conditions $D_{\alpha}\bar \Lambda=0$ and $\bar D_{\alpha }\Lambda=0$. Moreover, the higher-derivative operator in (\ref{puregauge}) was chosen to be linear in $\bar D^\alpha D_\alpha$ and $D^\alpha \bar D^2D_\alpha$ due to the identities:
\bea
\label{usefulid1}
(\bar D^\alpha D_\alpha)^n&=&\Box^{\frac{n-1}{2}}\bar D^\alpha D_\alpha \ , \ n=2l-1,\\
\label{usefulid2}
(\bar D^\alpha D_\alpha)^n&=&-\Box^{\frac{n}{2}-1}D^\alpha\bar D^2 D_\alpha \ , \ n=2l,\\
\label{usefulid3}
(D^\alpha \bar D^2D_\alpha)^n&=&(-1)^{n+1}\Box^{n-1}D^\alpha \bar D^2D_\alpha \ , \ n=1,2,3,\ldots \ ,
\eea
where $l=1,2,3,\ldots$.

We can add to (\ref{puregauge}) the following gauge-fixing term:
\bea
\label{gaugefix}
S_{GF}=-\frac{1}{2\alpha}\int d^3xd^4\theta\, V\{D^2,\bar D^2\}V \ .
\eea
Of course, we could have used a gauge-fixing term more sophisticated involving higher derivatives like in \cite{Our1}, however, we will use (\ref{gaugefix}) for convenience. Besides, we know that $\delta V=i(\bar\Lambda-\Lambda)$ is an Abelian symmetry, and therefore the ghosts completely decouple.

Now, let us consider the matter sector. We will make two assumptions in order to simplify the model involving the matter superfields. First, we will demand that the matter action does not contain terms with higher derivatives. Second, we will not consider self-couplings involving only $\Phi$ or $\bar\Phi$ superfields. Having made these assumptions, the most generic matter action is given by
\bea
\label{matter}
S_{M}=\int d^3xd^4\theta K(\bar\Phi,\Phi) \ ,
\eea
where $K(\bar\Phi,\Phi)$ is the tree-level KEP.

In order to couple (\ref{matter}) to the gauge superfield, the function $K(\bar\Phi,\Phi)$ must firstly be invariant under the global transformation $\delta\Phi=i\lambda\Phi$, and $\delta\bar\Phi=-i\lambda\bar\Phi$. It follows that $K(\bar\Phi,\Phi)$ must satisfy the constraint
\bea
\label{constraint}
\bar\Phi\frac{\partial K(\bar\Phi,\Phi)}{\partial\bar\Phi}=\frac{\partial K(\bar\Phi,\Phi)}{\partial \Phi}\Phi \ .
\eea
In particular this constraint is satisfied if $K{(\bar\Phi,\Phi)}$ is a function of $\bar\Phi\Phi$.

Now, we can introduce the gauge superfield $V$ in (\ref{matter}) to obtain
\bea
\label{mattercoupled}
S_{M}=\frac{1}{2}\int d^3xd^4\theta[K(\bar\Phi e^{2gV},\Phi)+K(\bar\Phi, e^{2gV}\Phi)] \ ,
\eea
which is invariant under local transformations $\delta\Phi=i\Lambda(z)\Phi$, $\delta\bar\Phi=-i\bar\Lambda(z)\bar\Phi$, and $\delta V=i(\bar\Lambda-\Lambda)$.

Finally, the generic higher-derivative $\mathcal{N}=2$, $d=3$ gauge theory that we will study in this work follows from (\ref{puregauge}), (\ref{gaugefix}), and (\ref{mattercoupled}):
\bea
\label{generictheory}
S&=&\frac{1}{2}\int d^3xd^4\theta\big\{ V[f(\Box)\bar D^\alpha D_\alpha+h(\Box)D^\alpha \bar D^2D_\alpha-\frac{1}{\alpha}\{D^2,\bar D^2\}]V+K(\bar\Phi e^{2gV},\Phi)\nonumber\\
&&+K(\bar\Phi, e^{2gV}\Phi)\big\} \ .
\eea
The standard method of calculating the effective action is based on the methodology of the loop expansion \cite{BO}. To do this, we make a shift $\Phi\rightarrow\Phi+\phi$ in the superfield $\Phi$ (together with the analogous shift for the $\bar\Phi$), where now $\Phi$ is a background (super)field and $\phi$ is a quantum one. We assume that the gauge field $V$ is quantum. In order to calculate the effective action at the one-loop level, we have to keep only the quadratic terms in the quantum fluctuations $\phi$, $\bar\phi$, and $V$ . By using this prescription, we get from (\ref{generictheory})
\bea
\label{backaction1}
S_2[\bar\Phi,\Phi;\bar\phi,\phi,V]&=&\frac{1}{2}\int d^3xd^4\theta\big\{ V[f(\Box)\bar D^\alpha D_\alpha+h(\Box)D^\alpha \bar D^2D_\alpha-\frac{1}{\alpha}\{D^2,\bar D^2\}]V\nonumber\\
&&+\frac{(2g)^2}{2}(K_{\bar\Phi}\bar\Phi+K_{\Phi}\Phi+K_{\bar\Phi\bar\Phi}\bar\Phi^2+K_{\Phi\Phi}\Phi^2)V^2+2g(K_{\bar\Phi}+K_{\bar\Phi\bar\Phi}\bar\Phi\nonumber\\
&&+K_{\bar\Phi\Phi}\Phi)\bar\phi V+2g(K_{\Phi}+K_{\Phi\Phi}\Phi+K_{\bar\Phi\Phi}\bar\Phi)V\phi+2K_{\bar\Phi\Phi}\bar\phi\phi\big\} \ ,
\eea
where the derivatives of the background superfields were omitted due to our interest only in the KEP \cite{BuKu}. By differentiating the constraint (\ref{constraint}) we obtain new identities which can be used to simplify (\ref{backaction1}), then we get
\bea
\label{backaction2}
&&S_2[\bar\Phi,\Phi;\bar\phi,\phi,V]=S_q+S_{int} \ ,\\
&&S_q=\frac{1}{2}\int d^3xd^4\theta\big\{ V[f(\Box)\bar D^\alpha D_\alpha-\Box h(\Box)\Pi_{1/2}-\frac{1}{\alpha}\Box\Pi_0]V+2K_{\bar\Phi\Phi}\bar\phi\phi\big\} \ ,\\
&&S_{int}=\frac{1}{2}\int d^3xd^4\theta\big\{(2g)^2K_{\bar\Phi\Phi}\bar\Phi\Phi V^2+2(2g)K_{\bar\Phi\Phi}\Phi\bar\phi V+2(2g)K_{\bar\Phi\Phi}\bar\Phi V\phi\big\} \ ,
\eea
where we used the projection operators $\Pi_{1/2}\equiv-\Box^{-1}D^\alpha\bar D^2D_\alpha$ and $\Pi_{0}\equiv\Box^{-1}\{D^2,\bar D^2\}$, which together with the operator $\bar D^\alpha D_\alpha$ satisfy the properties
\bea
\label{prop1}
\Pi_{1/2}^2=\Pi_{1/2} \ , & \Pi_{0}^2=\Pi_{0} \ , & (\bar D^\alpha D_\alpha)^2=\Box\Pi_{1/2} \ ,\\
\label{prop2}
\Pi_{1/2}\Pi_{0}=0 \ \ , & \Pi_{0}\bar D^\alpha D_\alpha=0 \ \ , & \Pi_{1/2}\bar D^\alpha D_\alpha=\bar D^\alpha D_\alpha \ .
\eea
These properties can be used to deduce the identities (\ref{usefulid1}-\ref{usefulid3}). Moreover, we can use them to extract the propagators from $S_q$. Thus, in momentum space,  we obtain
\bea
\label{propagator1}
\langle V(1)V(2)\rangle&=&\big[X(p^2)\bar D^\alpha D_\alpha+Y(p^2)\Pi_{1/2}-\frac{\alpha}{p^2}\Pi_0\big]_1\delta_{12} \ ,\\
\label{propagator2}
\langle \bar\phi(1)\phi(2)\rangle &=& \Big(\frac{1}{K_{\bar\Phi\Phi}p^2}\Big)_1 \delta_{12} \ ,
\eea
where
\bea
\label{xy}
X(p^2)=\frac{f(-p^2)}{p^2\big[p^2h^2(-p^2)+f^2(-p^2)\big]} \ \ \ \text{and} \ \ \  Y(p^2)=-\frac{h(-p^2)}{p^2h^2(-p^2)+f^2(-p^2)} \ .
\eea
These propagators will be used for the one-loop calculations.

\section{One-loop calculations}

Let us start the calculations of the one-loop supergraphs contributing to the KEP. At the one-loop order, we will have two types of contributions. In the first, all diagrams involve only the gauge superfield propagators $\langle V(1)V(2)\rangle$ in the internal lines connecting the vertices $(2g)^2K_{\bar\Phi\Phi}\bar\Phi\Phi V^2$. Such supergraphs exhibit structures given at Fig. 1.

\begin{figure}[!h]
\begin{center}
\includegraphics[angle=0,scale=0.40]{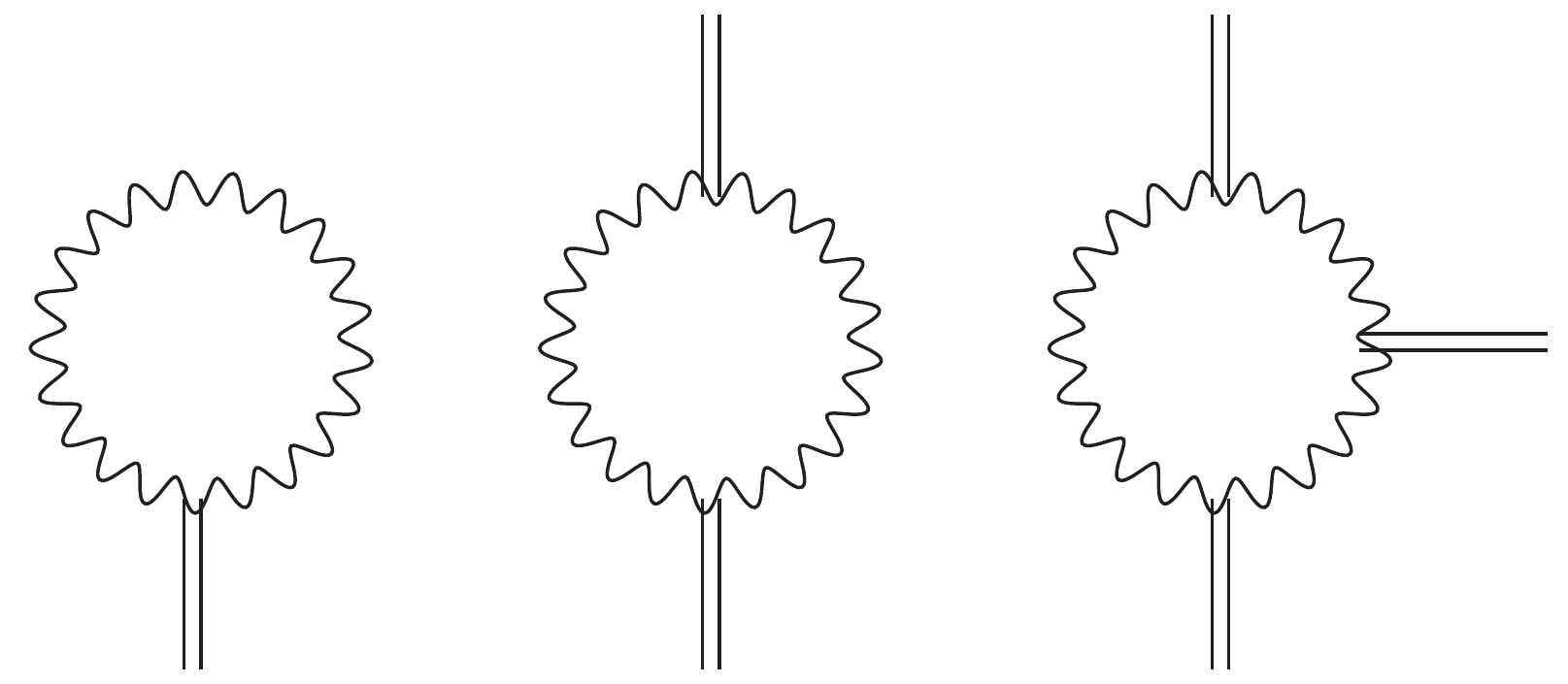}
\end{center}
\caption{One-loop supergraphs in a gauge sector.}
\end{figure}

We can compute all the contributions by noting that each supergraph above is formed by $n$ "subgraphs" like these ones given by Fig. 2.

\begin{figure}[!h]
\begin{center}
\includegraphics[angle=0,scale=0.60]{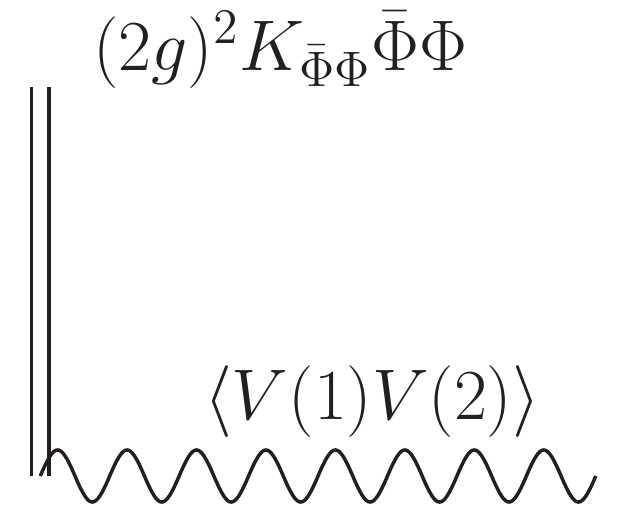}
\end{center}
\caption{A typical vertex in one-loop supergraphs in gauge sector.}
\end{figure}

 The contribution of this fragment is 
\bea
Q_{12}=[(2g)^2K_{\bar\Phi\Phi}\bar\Phi\Phi]_1\big(X\bar D^\alpha D_\alpha+Y\Pi_{1/2}-\frac{\alpha}{p^2}\Pi_0\big)_1\delta_{12} \ .
\eea
It follows from the result above that the contribution of a supergraph formed by $n$ fragments is given by
\bea
I_n&=&\int d^3x\frac{1}{2n}\int d^4\theta_1d^4\theta_2\ldots d^4\theta_{n}\int \frac{d^3p}{(2\pi)^3}Q_{12}Q_{23}\ldots Q_{n-1,n}Q_{n,1} \nonumber\\
&=&\int d^3x\frac{1}{2n}\int d^4\theta_1d^4\theta_2\ldots d^4\theta_{n}\int \frac{d^3p}{(2\pi)^3}[(2g)^2K_{\bar\Phi\Phi}\bar\Phi\Phi]_1\big(X\bar D^\alpha D_\alpha+Y\Pi_{1/2}\nonumber\\
&-&\frac{\alpha}{p^2}\Pi_0\big)_1\delta_{12} \ [(2g)^2K_{\bar\Phi\Phi}\bar\Phi\Phi]_2\big(X\bar D^\alpha D_\alpha+Y\Pi_{1/2}-\frac{\alpha}{p^2}\Pi_0\big)_2\delta_{23}\ldots\nonumber\\
&\times&[(2g)^2K_{\bar\Phi\Phi}\bar\Phi\Phi]_n\big(X\bar D^\alpha D_\alpha+Y\Pi_{1/2}-\frac{\alpha}{p^2}\Pi_0\big)_n\delta_{n,1} \ ,
\eea
where $2n$ is a symmetry factor. Such a contribution takes into account the Taylor series expansion coefficients of the effective action, the usual symmetry factor of each supergraph, and the number of topologically distinct supergraphs \cite{Hat}. The external momenta must  be taken to be zero in the calculation of the effective potential.

We can integrate by parts the expression $I_n$ and discard terms involving covariant derivatives of $\bar\Phi$ and $\Phi$ to get
\bea
\label{supergraphs}
I_n=\int d^3xd^4\theta\int \frac{d^3p}{(2\pi)^3}\frac{1}{2n}[(2g)^2K_{\bar\Phi\Phi}\bar\Phi\Phi]^n\big(X\bar D^\alpha D_\alpha+Y\Pi_{1/2}-\frac{\alpha}{p^2}\Pi_0\big)^n\delta_{\theta\theta^{\prime}}|_{\theta=\theta^{\prime}}.
\eea
The effective action is given by the sum of all supergraphs $I_n$
\bea
\Gamma^{(1)}_1&=&\sum_{n=1}^{\infty}I_n=\int d^3xd^4\theta\int \frac{d^3p}{(2\pi)^3}\sum_{n=1}^{\infty}\frac{1}{2n}[(2g)^2K_{\bar\Phi\Phi}\bar\Phi\Phi]^n\big[(X\bar D^\alpha D_\alpha+Y\Pi_{1/2})^n\delta_{\theta\theta^{\prime}}|_{\theta=\theta^{\prime}}\nonumber\\
&-&\frac{2}{p^2}\Big(-\frac{\alpha}{p^2}\Big)^n\big] \ ,
\eea
where we used (\ref{prop1}), (\ref{prop2}), and  the fact that $\Pi_0\delta_{\theta\theta^{\prime}}|_{\theta=\theta^{\prime}}=-2/p^2$. Summing over all $n$ we get
\bea
\Gamma^{(1)}_1&=&\int d^3xd^4\theta\int \frac{d^3p}{(2\pi)^3}\bigg\{-\frac{1}{2}\ln\Big[1-(2g)^2K_{\bar\Phi\Phi}\bar\Phi\Phi(X\bar D^\alpha D_\alpha+Y\Pi_{1/2})\Big]\delta_{\theta\theta^{\prime}}|_{\theta=\theta^{\prime}}\nonumber\\
&+&\frac{1}{p^2}\ln\bigg[1+\frac{\alpha(2g)^2K_{\bar\Phi\Phi}\bar\Phi\Phi}{p^2}\bigg]\bigg\} \ .
\eea
The first logarithm term can be splitted in two parts, then
\bea
\Gamma^{(1)}_1&=&\int d^3xd^4\theta\int \frac{d^3p}{(2\pi)^3}\bigg\{-\frac{1}{2}\ln\Big[1-\frac{(2g)^2K_{\bar\Phi\Phi}\bar\Phi\Phi X}{1-(2g)^2K_{\bar\Phi\Phi}\bar\Phi\Phi Y}\bar D^\alpha D_\alpha\Big]\delta_{\theta\theta^{\prime}}|_{\theta=\theta^{\prime}}\nonumber\\
&-&\frac{1}{2}\ln\Big[1-(2g)^2K_{\bar\Phi\Phi}\bar\Phi\Phi Y\Pi_{1/2}\Big]\delta_{\theta\theta^{\prime}}|_{\theta=\theta^{\prime}}+\frac{1}{p^2}\ln\bigg[1+\frac{\alpha(2g)^2K_{\bar\Phi\Phi}\bar\Phi\Phi}{p^2}\bigg]\bigg\} \ .
\eea
Finally, we expand in Taylor series the first two logarithms and use (\ref{usefulid1}-\ref{usefulid3}), (\ref{prop1}-\ref{prop2}), and $\Pi_{1/2}\delta_{\theta\theta^{\prime}}|_{\theta=\theta^{\prime}}=2/p^2$ to obtain
\bea
\label{part1}
\Gamma^{(1)}_1&=&\int d^3xd^4\theta\int \frac{d^3p}{(2\pi)^3}\frac{1}{p^2}\bigg\{-\frac{1}{2}\ln\Big[1+p^2\Big(\frac{(2g)^2K_{\bar\Phi\Phi}\bar\Phi\Phi X}{1-(2g)^2K_{\bar\Phi\Phi}\bar\Phi\Phi Y}\Big)^2\Big]\nonumber\\
&-&\ln\Big[1-(2g)^2K_{\bar\Phi\Phi}\bar\Phi\Phi Y\Big]+\ln\bigg[1+\frac{\alpha(2g)^2K_{\bar\Phi\Phi}\bar\Phi\Phi}{p^2}\bigg]\bigg\} \ .
\eea
Let us proceed the calculation of the second type of one-loop supergraphs, which involve the gauge and matter superfield propagators in the internal lines connecting the vertices $(2g)K_{\bar\Phi\Phi}\Phi\bar\phi V$ and $(2g)K_{\bar\Phi\Phi}\bar\Phi V\phi$. Such supergraphs exhibit the structure shown in Fig. 3.

\begin{figure}[!h]
\begin{center}
\includegraphics[angle=0,scale=0.45]{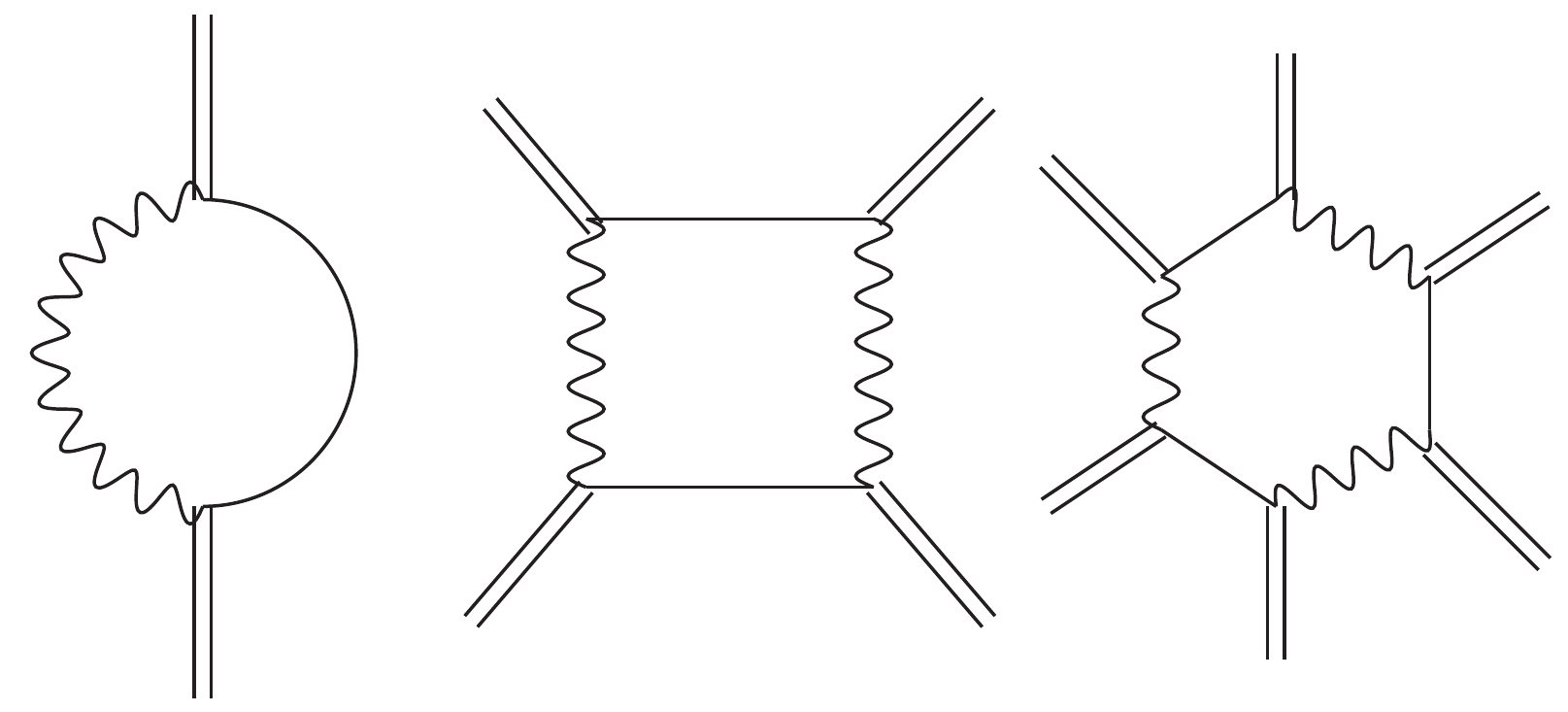}
\end{center}
\caption{One-loop supergraphs in a mixed sector.}
\end{figure}

\begin{figure}[!h]
\begin{center}
\includegraphics[angle=0,scale=0.60]{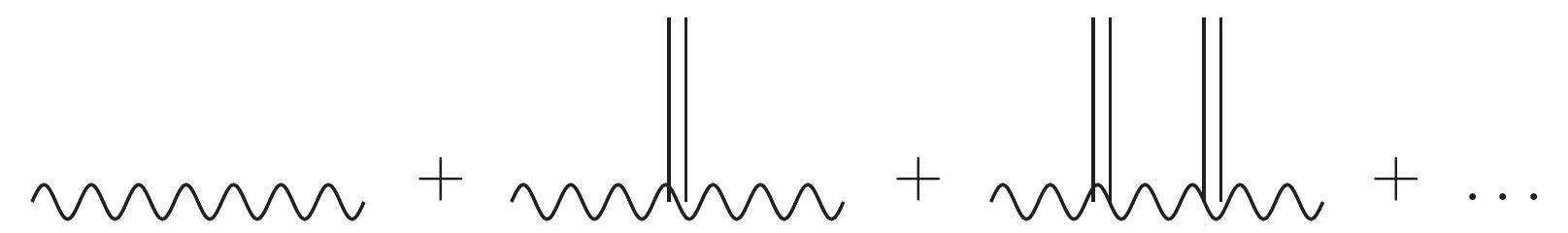}
\end{center}
\caption{Dressed propagator.}
\end{figure}

It is worth to point out that we can insert an arbitrary number of vertices $(2g)^2K_{\bar\Phi\Phi}\bar\Phi\Phi V^2$ into the gauge propagators. Therefore, we should firstly introduce a "dressed" propagator. In this propagator, the summation over all vertices $(2g)^2K_{\bar\Phi\Phi}V^2$ is performed (see Fig. 4). As a result, this dressed propagator is equal to
\bea
\langle V(1)V(2)\rangle_D&=&\langle V(1)V(2)\rangle+\int d^4\theta_3\langle V(1)V(3)\rangle[(2g)^2K_{\bar\Phi\Phi}\bar\Phi\Phi]_3\langle V(3)V(2)\rangle\nonumber\\
&+&\int d^4\theta_3d^4\theta_4\langle V(1)V(3)\rangle[(2g)^2K_{\bar\Phi\Phi}\bar\Phi\Phi]_3\langle V(3)V(4)\rangle[(2g)^2K_{\bar\Phi\Phi}\bar\Phi\Phi]_4\nonumber\\
&\times&\langle V(4)V(2)\rangle+\ldots \ .
\eea
By using (\ref{propagator1}) and integrating by parts, we arrive at
\bea
\label{dressedpropagator}
\langle V(1)V(2)\rangle_D=\sum_{n=0}^\infty[(2g)^2K_{\bar\Phi\Phi}\bar\Phi\Phi]^n_1\big[(X\bar D^\alpha D_\alpha+Y\Pi_{1/2})^{n+1}+\Big(-\frac{\alpha}{p^2}\Big)^{n+1}\Pi_{0}\big]_1\delta_{12} \ .
\eea
As before, we can compute all the contributions by noting that each supergraph above (Fig. 3) is formed by $n$ subgraphs, like those depicted in Fig.~5 and Fig. 6. Since both subgraphs, Figs. 5 and 6, provide the same contribution, we just need to calculate the one in the Fig. 5. This subgraph yields the contribution $(\Pi_-\equiv-\bar D^2D^2/p^2)$
\bea
R_{13}&=&\int d^4\theta_2[(2g)K_{\bar\Phi\Phi}\Phi]_1\big\{\sum_{n=0}^\infty[(2g)^2K_{\bar\Phi\Phi}\bar\Phi\Phi]^n_1\big[(X\bar D^\alpha D_\alpha+Y\Pi_{1/2})^{n+1}\nonumber\\
&+&\Big(-\frac{\alpha}{p^2}\Big)^{n+1}\Pi_{0}\big]_1\delta_{12}\big\}[(2g)K_{\bar\Phi\Phi}\bar\Phi]_2\Big[-\Big(\frac{\Pi_-}{K_{\bar\Phi\Phi}}\Big)_2 \delta_{23}\Big]\nonumber\\
&=&-\sum_{n=0}^\infty[(2g)^2K_{\bar\Phi\Phi}\bar\Phi\Phi]^{n+1}_1\Big(-\frac{\alpha}{p^2}\Big)^{n+1}\Big(\Pi_-\Big)_1\delta_{13} \ .
\eea
By summing up, we arrive at
\bea
R_{13}=\bigg(\frac{(2g)^2\alpha K_{\bar\Phi\Phi}\bar\Phi\Phi}{p^2+(2g)^2\alpha K_{\bar\Phi\Phi}\bar\Phi\Phi}\Pi_-\bigg)_1\delta_{13} \ .
\eea

\begin{figure}[!h]
\begin{center}
\includegraphics[angle=0,scale=0.60]{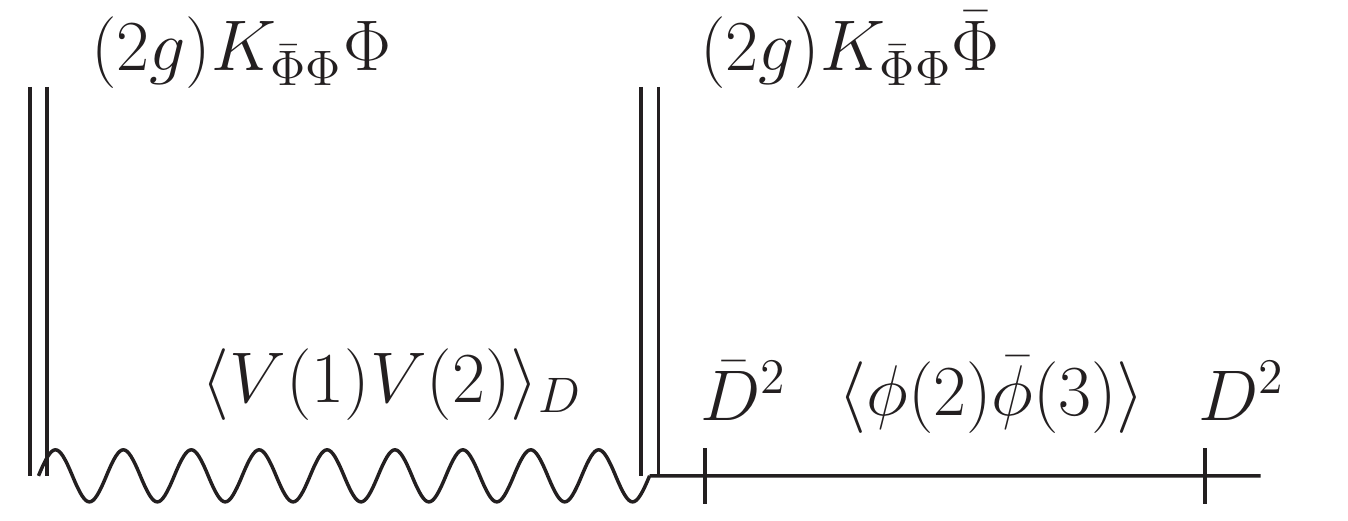}
\end{center}
\caption{A typical link in one-loop supergraphs in mixed sector.}
\end{figure}

\begin{figure}[!h]
\begin{center}
\includegraphics[angle=0,scale=0.60]{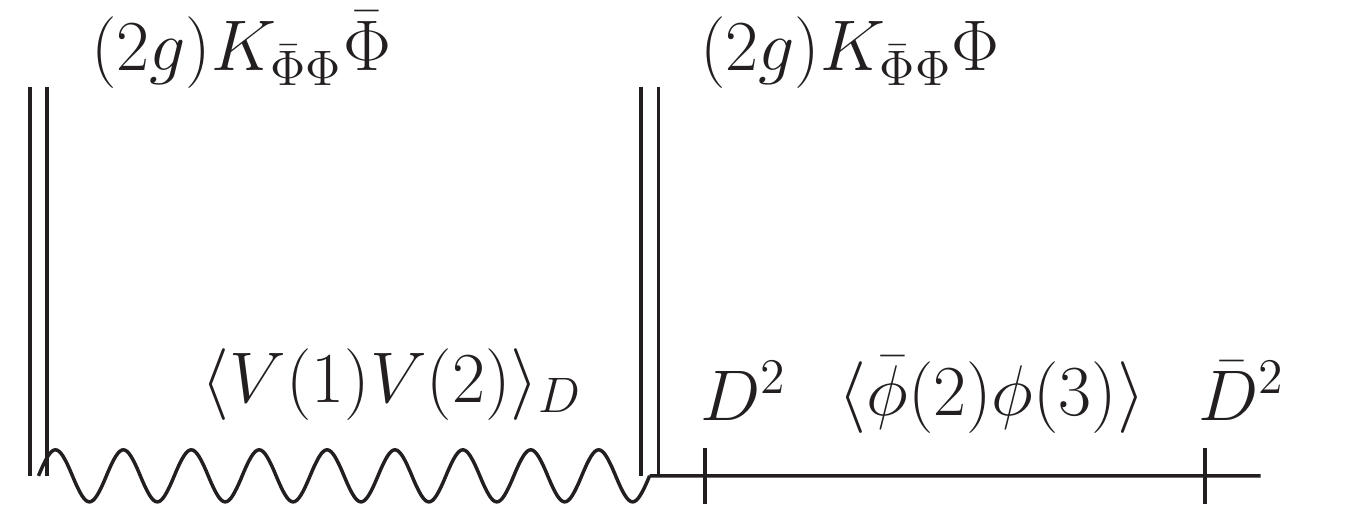}
\end{center}
\caption{Another typical link in one-loop supergraphs in mixed sector.}
\end{figure}

It follows from the result above that the contribution of a supergraph formed by $n$ subgraphs is given by
\bea
J_n&=&\int d^3x\frac{1}{2n}\int d^4\theta_1d^4\theta_3\ldots d^4\theta_{2n-1}\int \frac{d^3p}{(2\pi)^3}R_{13}R_{35}\ldots R_{2n-3,2n-1} R_{2n-1,1} \nonumber\\
&=&\int d^3x\frac{1}{2n}\int d^4\theta_1d^4\theta_3d^4\theta_5\ldots d^4\theta_{2n-1}\int \frac{d^3p}{(2\pi)^3}\Big[\bigg(\frac{(2g)^2\alpha K_{\bar\Phi\Phi}\bar\Phi\Phi}{p^2+(2g)^2\alpha K_{\bar\Phi\Phi}\bar\Phi\Phi}\Pi_-\bigg)_1\delta_{13}\Big]\nonumber\\
&\times&\Big[\bigg(\frac{(2g)^2\alpha K_{\bar\Phi\Phi}\bar\Phi\Phi}{p^2+(2g)^2\alpha K_{\bar\Phi\Phi}\bar\Phi\Phi}\Pi_-\bigg)_3\delta_{35}\Big]\ldots
\Big[\bigg(\frac{(2g)^2\alpha K_{\bar\Phi\Phi}\bar\Phi\Phi}{p^2+(2g)^2\alpha K_{\bar\Phi\Phi}\bar\Phi\Phi}\Pi_-\bigg)_{2n-1}\delta_{2n-1,1}\Big] \nonumber\\
&=&\int d^3xd^4\theta\frac{1}{2n}\int \frac{d^3p}{(2\pi)^3}\bigg(\frac{(2g)^2\alpha K_{\bar\Phi\Phi}\bar\Phi\Phi}{p^2+(2g)^2\alpha K_{\bar\Phi\Phi}\bar\Phi\Phi}\bigg)^n\Pi_-\delta_{\theta\theta^{\prime}}|_{\theta=\theta^{\prime}}\ .
\eea
By using $\Pi_-\delta_{\theta\theta^{\prime}}|_{\theta=\theta^{\prime}}=-1/p^2$, we get the effective action
\bea
\label{part2}
\Gamma^{(1)}_2=2\sum_{n=0}^{\infty}J_n=-\int d^3xd^4\theta\frac{1}{p^2}\ln\bigg[1+\frac{\alpha(2g)^2K_{\bar\Phi\Phi}\bar\Phi\Phi}{p^2}\bigg] \ .
\eea
It is worth to point out that the contribution (\ref{part2}) cancels the dependence of (\ref{part1}) on the gauge parameter $\alpha$. By summing (\ref{part1}) to (\ref{part2}) we obtain the total one-loop effective action
\bea
\label{totalEA}
\Gamma^{(1)}[\bar\Phi,\Phi]&=&\int d^3xd^4\theta\int \frac{d^3p}{(2\pi)^3}\frac{1}{p^2}\bigg\{-\frac{1}{2}\ln\Big[1+p^2\Big(\frac{(2g)^2K_{\bar\Phi\Phi}\bar\Phi\Phi X}{1-(2g)^2K_{\bar\Phi\Phi}\bar\Phi\Phi Y}\Big)^2\Big]\nonumber\\
&-&\ln\Big[1-(2g)^2K_{\bar\Phi\Phi}\bar\Phi\Phi Y\Big]\bigg\} \ .
\eea
Finally, we arrive to the following result for the KEP (as usual, the corresponding effective
action can be restored from the relation $\Gamma^{(1)}=\int d^3xd^4\theta K^{(1)})$:
\bea
\label{totalEP}
K^{(1)}(\bar\Phi,\Phi)&=&\int \frac{d^3p}{(2\pi)^3}\frac{1}{p^2}\bigg\{-\frac{1}{2}\ln\Big[1+p^2\Big(\frac{(2g)^2K_{\bar\Phi\Phi}\bar\Phi\Phi X}{1-(2g)^2K_{\bar\Phi\Phi}\bar\Phi\Phi Y}\Big)^2\Big]\nonumber\\
&-&\ln\Big[1-(2g)^2K_{\bar\Phi\Phi}\bar\Phi\Phi Y\Big]\bigg\} \ ,
\eea
where, $X$ and $Y$ are given by (\ref{xy}). Moreover, notice that (\ref{totalEP}) is independent of the gauge parameter for any choice of $K(\bar\Phi e^{2gV}\Phi)$, $f(-p^2)$, and $h(-p^2)$.

The result (\ref{totalEP}) is rather generic. Therefore, in order to proceed with the calculation and solve explicitly the integral above,  we have to specify the operators $f(\Box)$ and $h(\Box)$ in (\ref{xy}). So, let us consider two characteristic examples where the final result is expressed in closed form and in terms of elementary functions.

As our first example, let us take $f(\Box)=\xi_f(-\Box)^n$ and $h(\Box)=0$ in (\ref{xy}), where $\xi_f$ is a parameter with a nontrivial mass dimension $[\xi_f]=[M]^{-2n+1}$, $\xi_f>0$, and $n$ is a non-negative integer. This choice corresponds to a higher-derivative Chern-Simons theory (see (\ref{puregauge})). It follows from (\ref{totalEP}) that
\bea
\label{CSEP1}
K^{(1)}_{HCS}(\bar\Phi,\Phi)=-\frac{1}{2}\int \frac{d^3p}{(2\pi)^3}\frac{1}{p^2}\ln\Big[1+\frac{1}{(p^2)^{2n+1}}\Big(\frac{(2g)^2K_{\bar\Phi\Phi}\bar\Phi\Phi}{\xi_f}\Big)^2\Big] \ ,
\eea
whose solution is given by
\bea
\label{CSEP2}
K^{(1)}_{HCS}(\bar\Phi,\Phi)=-\frac{1}{4\pi}\csc{\Big[\frac{\pi}{2(2n+1)}\Big]}\Big(\frac{(2g)^2K_{\bar\Phi\Phi}\bar\Phi\Phi}{\xi_f}\Big)^{\frac{1}{2n+1}} \ .
\eea
The second example is $f(\Box)=0$ and $h(\Box)=\xi_h(-\Box)^n$ in (\ref{xy}), where $[\xi_h]=[M]^{-2n}$, $\xi_h>0$. This choice corresponds to a higher-derivative Maxwell theory. It follows from (\ref{totalEP}) that
\bea
\label{MEP1}
K^{(1)}_{HQED}(\bar\Phi,\Phi)&=&-\int \frac{d^3p}{(2\pi)^3}\frac{1}{p^2}\ln\Big[1+\frac{1}{(p^2)^{n+1}}\Big(\frac{(2g)^2K_{\bar\Phi\Phi}\bar\Phi\Phi}{\xi_h}\Big)\Big] \ ,
\eea
whose solution is given by
\bea
\label{MEP2}
K^{(1)}_{HQED}(\bar\Phi,\Phi)=-\frac{1}{2\pi}\csc{\Big[\frac{\pi}{2(n+1)}\Big]}\Big(\frac{(2g)^2K_{\bar\Phi\Phi}\bar\Phi\Phi}{\xi_h}\Big)^{\frac{1}{2(n+1)}} \ .
\eea
We notice that the one-loop corrections for the KEPs, namely (\ref{CSEP2},\ref{MEP2}), are finite and do not need any renormalization. Moreover, these results are universal, valid for any form of the potential $K(\bar\Phi e^{2gV}\Phi)$. We also notice that the functional structure of (\ref{CSEP2}) and (\ref{MEP2}) does not involve any logarithm-like dependence, which is usually found in four-dimensional theories.
We observe that, up to constants, $K^{(1)}_{HQED}(\bar\Phi,\Phi)$ given in (\ref{MEP2})  is the same as in  the $\mathcal{N}=1$ case derived in \cite{Our1}.
 Additionally, in \cite{Our1} it  was shown that the one-loop KEP vanishes for the $\mathcal{N}=1$, $d=3$ QED coupled to non-self-interacting matter; we see from (\ref{MEP2}) that  it is not the case for $\mathcal{N}=2$, $d=3$ QED coupled to non-self-interacting matter.

\section{Summary}

We have calculated the one-loop effective potential for the ${\cal N}=2$ supersymmetric three-dimensional higher-derivative theories. Our calculation was based on a formalism allowing to maintain the ${\cal N}=2$ supersymmetry at all steps of the calculations. Discussing the properties of the result, we should emphasize, first, its finiteness which is a natural consequence of the presence of  higher derivatives, second, the similarity of its form to the results obtained earlier for the four-dimensional ${\cal N}=1$ supersymmetric theories \cite{ourhigh}. Also, contrarily to the case of the ${\cal N}=1$, $d=3$, supersymmetric QED \cite{Our1}, in our theory the one-loop k\"{a}hlerian effective potential does not vanish. However, the one-loop  $\mathcal{N}=1$ and $\mathcal{N}=2$ k\"{a}hlerian effective potentials for the supersymmetric Chern-Simons theories in  $d=3$ display
similar structures.

\vspace*{4mm}

{\bf APPENDIX}

\vspace*{2mm}

Let us briefly describe the $\mathcal{N}=2$ supersymmetry algebra used in this paper. Here we work within the three-dimensional Minkowski space, so, we choose the gamma matrices as $(\gamma^{\mu})^{\alpha}_{\phantom{\alpha}\beta}=(\sigma^2,i\sigma^1,i\sigma^3)$, which satisfy the Clifford algebra $\{\gamma^\mu,\gamma^\nu\}=-2\eta^{\mu\nu}$, with $\eta^{\mu\nu}=\text{diag}(-1,1,1)$. We raise and lower spinor indices with the matrix $C_{\alpha\beta}=\sigma^2$, so that $C^{12}=-C_{12}=i$, and $\psi^{\alpha}=C^{\alpha\beta}\psi_{\beta}$, $\psi_{\beta}=\psi^{\alpha}C_{\alpha\beta}$, $\psi^2=\frac{1}{2}C_{\beta\alpha}\psi^\alpha\psi^\beta$. It follows from $(\gamma^\mu)_{\alpha\beta}=C_{\gamma\alpha}{(\gamma^\mu)^\gamma}_\beta$ and $(\gamma_\mu)^{\alpha\beta}=\eta_{\mu\nu} C^{\beta\lambda}{(\gamma^\nu)^\alpha}_\lambda$ that
\bea
(\gamma^\mu)_{\alpha\beta}=\{-\hat{I},-\sigma^3,\sigma^1\} \ , \ (\gamma_\mu)^{\alpha\beta}=\{-\hat{I},-\sigma^3,\sigma^1\} \ .
\eea
From these equations, we get
\bea
(\gamma^\mu)_{\alpha\beta}(\gamma_\nu)^{\alpha\beta}=2{\delta^\mu}_\nu \ , \ (\gamma^\mu)_{\alpha\beta}(\gamma_\mu)^{\gamma\delta}=({\delta_\alpha}^\gamma{\delta_\beta}^\delta
+{\delta_\alpha}^\delta{\delta_\beta}^\gamma) \ .
\eea
Therefore, we can use the gamma matrices to map the components of 3-vectors into $2\times2$ symmetric (hermitian) matrices by means of the definitions
\bea
&&\text{For fields}: \ \ V^{\alpha\beta}=\frac{1}{\sqrt{2}}{(\gamma_\mu)}^{\alpha\beta}V^\mu \ , \ V^\mu=\frac{1}{\sqrt{2}}{(\gamma^\mu)}_{\alpha\beta}V^{\alpha\beta} \ ; \\
&&\text{For derivatives}: \ \ \partial_{\alpha\beta}={(\gamma^\mu)}_{\alpha\beta}\partial_\mu \ , \ \partial_\mu=\frac{1}{2}{(\gamma_\mu)}^{\alpha\beta}\partial_{\alpha\beta} \ ;\\
&&\text{For coordinates}: \ \ x^{\alpha\beta}=\frac{1}{2}{(\gamma_\mu)}^{\alpha\beta}x^\mu \ , \ x^\mu={(\gamma^\mu)}_{\alpha\beta}x^{\alpha\beta} \ .
\eea
The $\mathcal{N}=2$, $d=3$ supersymmetry algebra is
\bea
\{Q^i_\alpha,Q^j_\beta\}=2\delta^{ij}P_{\alpha\beta} \, \ (i,j=1,2) \ ,
\eea
where $P_{\alpha\beta}=i\partial_{\alpha\beta}$.
However, it is convenient to go over to a complex representation by defining
\bea
Q_\alpha=\frac{1}{2}(Q^1_\alpha+iQ^2_\alpha) \ , \ \bar Q_\alpha=\frac{1}{2}(Q^1_\alpha-iQ^2_\alpha) \ ,
\eea
which  can  be  used  to  express  the  algebra  as,
\bea
\{Q_\alpha,\bar Q_\beta\}=P_{\alpha\beta} \ , \ \{Q_\alpha, Q_\beta\}=0 \ , \ \{\bar Q_\alpha, \bar Q_\beta\}=0 \ .
\eea
Notice that these conventions and definitions are the exact analogues of those ones used in \cite{SGRS}. In fact, the $\mathcal{N}=2$, $d=3$ superspace can be parametrized by the coordinates $z^M=(x^{\alpha\beta},\theta^\alpha,\bar\theta^\alpha)$, with $(\theta^\alpha)^*=\bar\theta^\alpha$, and the explicit forms of the  generators and covariant derivatives are given by
\bea
Q_\alpha=i(\partial_\alpha-\frac{1}{2}\bar\theta^{\beta}i\partial_{\alpha\beta}) \ &,& \ \bar Q_{\alpha}=i(\bar\partial_{\alpha}-\frac{1}{2} \theta^\beta i\partial_{\alpha\beta}),\\
D_\alpha=\partial_\alpha+\frac{1}{2}\bar\theta^{\beta}i\partial_{\alpha\beta} \ &,& \ \bar D_{\alpha}=\bar\partial_{\alpha}+\frac{1}{2} \theta^\beta i\partial_{\alpha\beta} \ .
\eea
We note that despite the derivatives $D_{\alpha}$ and $\bar{D}_{\beta}$ are independent, there is no chirality in this case since both types of the derivatives (and of the spinors) are transformed under the same (unique) spinor representation of the Lorentz group.

The (anti)commutation relations for the $D_{\alpha}$ and $\bar{D}_{\alpha}$ are rather similar to those ones for the four-dimensional supersymmetry \cite{SGRS}. Indeed, one has
\bea
&& \{D_{\alpha},\bar{D}_{\beta}\}=i\partial_{\alpha\beta};\quad\, \{D_{\alpha},D_{\beta}\}=\{\bar{D}_{\alpha},\bar{D}_{\beta}\}=0, \quad\, D_{\alpha}D^2=\bar{D}_{\alpha}\bar{D}^2=0; \nonumber\\
&& D^{\alpha}D_{\beta}=\delta^{\alpha}_{\beta}D^2; \quad\, \bar{D}^{\alpha}\bar{D}_{\beta}=\delta^{\alpha}_{\beta}\bar{D}^2; \quad\, [D^{\alpha},\bar{D}^2]=i\partial^{\alpha\beta}\bar{D}_{\beta}; \quad\, [\bar{D}^{\alpha},D^2]=i\partial^{\alpha\beta}D_{\beta};\nonumber\\
&& \bar{D}^2D^2\bar{D}^2=\Box \bar{D}^2; D^2\bar{D}^2D^2=\Box D^2.
\eea
These (anti)commutation relations can be used to prove the identities (\ref{usefulid1}-\ref{usefulid3}) and (\ref{prop1}-\ref{prop2}). Moreover, it is clear that the
 use of the derivatives satisfying these rules is no more difficult as  the use of the standard supercovariant derivatives either in three- or in four-dimensional case.

Finally, all quantum calculations were carried out using a Wick-rotated metric $\eta^{\mu\nu}=\text{diag}(+1,1,1)$.

\vspace{5mm}

{\bf Acknowledgments.}
This work was partially supported by Conselho Nacional de
Desenvolvimento Cient\'\i fico e Tecnol\'ogico (CNPq). A. Yu. P. has
been supported by the CNPq project No. 303438-2012/6. The work by F. S. Gama has been supported by the
CNPq process No. 141228/2011-3.

\end{document}